\documentstyle[aps,prl,floats,epsf,amsmath,amssymb]{revtex}

\begin{document}
\advance\textheight by 0.2in
\draft

\twocolumn[\hsize\textwidth\columnwidth\hsize\csname@twocolumnfalse%
\endcsname

\title{Power fluctuations in a driven damped chaotic pendulum}
\author{Yadin Y. Goldschmidt}
\address{Department of Physics and Astronomy, 
\\University of Pittsburgh, Pittsburgh PA 15260, U.S.A.}
\date{\today}
\maketitle

\begin{abstract}
In this paper we investigate the power fluctuations in a driven,
dampted pendulum. When the motion of the pendulum is chaotic, the
average power supplied by the driving force is equal to the average
dissipated power only for an infinite long time period. We measure the
fluctuations of the supplied power during a time equal to the period
of the driving force. Negative power fluctuations occur and we
estimate their probability. In a chaotic state
the histogram of the power distribution is broad and continuous although
bounded. For a value of the power not too close to the edge of the distribution the
Fluctuation Theorem of Gallavotti and Cohen is approximately satisfied.
\end{abstract}
]
Recently Gallavotti and Cohen (GC) \cite{GC} derived a fluctuation theorem (FT) for
chaotic systems. This theorem is concerned with the rate of entropy production
$\sigma_{\tau}$ averaged over an observation time $\tau$. It states that the
ratio of probabilities of having a given entropy production $\sigma_{\tau}$ to
that of having ($-\sigma_{\tau})$ is given by $\exp(\tau\sigma_{\tau})$. The
theorem was proven for thermostatted Hamiltonian systems which are driven by
external forces, under the condition that the system is sufficiently chaotic
(of the Anosov type \cite{gal}) and the number of degrees of freedom is very large. All
the systems considered had a certain form of time reversibility even though
they were dissipative. This was the result of the way the thermostat was
imposed. The fluctuation theorem is actually a generalization of the
fluctuation dissipation theorem (FDT) \ to systems away from equilibrium i.e.
driven systems in a steady state. Gallavotti \cite{gallavotti} has shown that
in the limit of vanishing driving force the FT reduces to the FDT.

More recently Kurchan \cite{kurchan} and subsequently Lebowitz and Spohn
\cite{LS} extended the FT to stochastic dynamics. Kurchan showed that the
FT\ holds for finite systems undergoing Langevin dynamics. The role of chaos
is replaced by the stochastic fluctuations. Lebowitz and Spohn generalized the
FT\ to general Markov processes. Particular role is played by systems
displaying a form of local detailed balance.

Our goal in this work was to test the FT for the case of a simple driven
dissipative chaotic system that does \textit{not} satisfy the conditions set
by GC. For example, this system is most likely not of the Anosov type \cite{york}, and it
is also not strictly thermostatted (energy is conserved only over a long
period of time). Thus, it is deterministic but it is not time reversible
because of the role of dissipation. This example may serve as a representative
that is more closely related to real life realizations. Surprisingly, (or not)
we find that approximately, the FT is satisfied, taking into account that we
have only three degrees of freedom, whereas the FT derived by GC should apply for
the case of a large number of degrees of freedom.

The system that we consider is that of a driven damped pendulum
\cite{marion,humieres,gwinn,baker}. The driving force is periodic. Varying the
strength of the driving force the pendulum can display simple periodic
behavior, a motion characterized by period doubling, tripling etc., or a fully
chaotic motion. There is a lot of literature on such a system. What we had in
mind was to consider the power bestowed on the system by the driving force
during a complete period of the latter (or during an integer multiple of the
period). When the motion is periodic, the value of the power bestowed during
any full period of the driving force is constant. But when the system is fully
chaotic, there is a broad distribution and occasionally the power bestowed is
negative. We are going to investigate numerically the power distribution, and
discuss the similarity to the GC Fluctuation Theorem. We are going to find
that the power distribution is actually a good fingerprint to determine if the
system is truly chaotic.

We consider a planar pendulum subject to the force of gravity, a damping force
resulting for instance from a viscous medium in which the pendulum is
immersed, and a sinusoidal driving force acting around the pivot point. The
equation of motion for the angle $\theta$ is \cite{marion}:%
\begin{equation}
\ddot{\theta}=-\frac{b}{m\ell^{2}}\dot{\theta}-\frac{g}{\ell}\sin\theta
+\frac{N_{d}}{m\ell^{2}}\cos\omega_{d}t.\label{pendulum}%
\end{equation}
Here $b$ is the damping coefficient and $N_{d}$ is the driving torque of
angular frequency $\omega_{d}$. We are going to switch to dimensionless
parameters. Denoting by $\omega_{0}^{2}=g/\ell$ and $t_{0}=1/\omega_{0}$, we
set $t^{\prime}=t/t_{0}$ and $\omega=\omega_{d}/\omega_{0}$. We also define%
\begin{equation}
x=\theta,\ \ c=\frac{b}{m\ell^{2}\omega_{0}},\ \ F=\frac{N_{d}}{mg\ell
}.\label{param}%
\end{equation}
The equation of motion can now be written as a set of two first order
differential equations (omitting the prime over the rescaled time variable):%
\begin{align}
\dot{x}(t) &  =y,\label{1storder}\\
\dot{y}(t) &  =-cy-\sin(x)+F\cos(\omega t).\label{de}%
\end{align}

We have solved these equations numerically using Mathematica. We have used the
values
\begin{equation}
c=0.05,\ \ \omega=0.7 \label{co}%
\end{equation}
and considered different values of the driving strength $F$ in the range
0-1.0. Defining%
\begin{equation}
\tau_{d}=\frac{2\pi}{\omega},
\end{equation}
to be the period of the driving force, we will be interested in evaluating the
quantity%
\begin{equation}
\left\langle P\right\rangle _{\tau,m}=\frac{F}{\tau}\int_{t_m}^{t_m+\tau
}dt\ y(t)\ \cos(\omega t), \label{power}%
\end{equation}
with%
\begin{equation}
\tau=n \tau_{d}, \ \ \ t_m=m \tau \label{tau}%
\end{equation}
and $n$ is a fixed integer. We will be particularly interested in the case $n=1$.
$\left\langle P\right\rangle _{\tau}$ is the average
power supplied to the system by the driving force during $n$ complete cycles
of the latter. 
In the following we will calculate $\left\langle P\right\rangle
_{\tau,m}$ along a phase-space trajectory of the system, starting
with m=100 to allow the transient to decay. We calculate consecutive
averages over time periods of duration $\tau$
and observe the fluctuations of the measured quantities.

If we multiply Eq.(\ref{de}) by $y$ we get
\begin{equation}
\frac{d}{dt}\left(  \frac{1}{2}y^{2}-\cos x\right)  =Fy\cos(\omega t)-cy^{2}.
\end{equation}
Thus averaging over an arbitrary time period $\tau$
\begin{equation}
\left\langle \frac{d}{dt}E\right\rangle _{\tau}=F\left\langle y\cos(\omega
t)\right\rangle _{\tau}-c\left\langle y^{2}\right\rangle _{\tau}, \label{avp}%
\end{equation}
where $E$ is the total energy of the pendulum (kinetic plus gravitational).
The first term on the right hand side of the equation represents the power
input from the driving force and the second term represents the power output
through dissipation with e.g. a viscous medium. The second term is always
positive, whereas there is no restriction about the sign of the first term. We
expect that in the limit of large $\tau$, the left hand side ($=(E(\tau
)-E(0))/\tau$) will approach zero, if the system is in a steady state, and
thus in that limit the two terms on the right should cancel each other. For a
simple harmonic oscillator, i.e. if the $\sin x$ in Eq.(\ref{de}) is replaced
by $x$, then after the transient term dies out the left hand side of Eq.
(\ref{avp}) is zero for any $\tau=n\ \tau_{d}$ and
\begin{equation}
\left\langle P\right\rangle _{\tau}=\frac{c}{2}\frac{F^{2}\omega^{2}%
}{(1-\omega^{2})^{2}+\omega^{2}c^{2}}. \label{sho}%
\end{equation}
It has a maximum value of \ $F^{2}/(2c)$ when $\omega=1$ (at resonance). This
result is of course not applicable for the pendulum where the non-linear
potential is present. We will see below that for the pendulum with fixed $c$
and $\omega$, $\left\langle P\right\rangle _{\tau}$ behaves very differently
as a function of $F$ than this simple quadratic form.

For small $n$, and in particular for $n=1$, we will be interested in the
fluctuations in $\left\langle P\right\rangle _{\tau}$, and in particular in
its histogram. We will see that $\left\langle P\right\rangle _{\tau}$ has a
non-vanishing probability to be negative in the chaotic regime, and we will
compare the probability of having $\left\langle P\right\rangle _{\tau}=a$, to
the probability of $\left\langle P\right\rangle _{\tau}=-a$, within the valid
range of $a$.

We should also emphasize here the difference between our system and a
thermostatted system considered in Ref. (\cite{GC}). For a thermostatted
system the dissipation is adjusted in such a way, by adjusting its velocity
dependence, that the energy of the system remain identically constant, not
only on the average over a long period of time, as is the case here.

Varying the strength of the applied force the pendulum can display a simple
periodic motion, a motion displaying periodic doubling, tripling, etc., or
truly chaotic motion. In Fig. 1 we show the average power supplied by the
driving force to the pendulum by using a time period of $\tau=400\ \tau_{d}%
$\ for the averaging (after the transient component has practically died). The
same results were obtained for the average power dissipated i.e.
$c\left\langle y^{2}\right\rangle _{\tau}$. It is evident that beyond the
initial segment F=0-0.4 which corresponds to periodic motion, the power curve is
not smooth but vary significantly from point to point. The big jump in the
power that occurs for $F\sim0.41$ is associated with the threshold for the
pendulum to go over the top. The error bars are associated with chaotic motion
when we do not have enough statistics to pinpoint the average power exactly.

\begin{figure}[t]
\centerline{\epsfysize 8cm \epsfbox{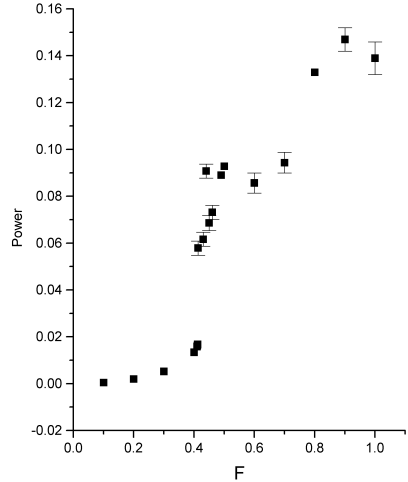}} \vspace{3mm}
\caption{The average power vs. the magnitude of the driving force. }  
\label{fpower}
\end{figure}

We now consider the histogram of the values of the supplied power during a
time period $\tau=\tau_{d}$. When the motion is periodic, the histogram
consists of a single peak. For example, for F=0.4 we get a single peak at
=0.0134. For the case of F=0.5 there is period tripling, i.e. the Poincar\'e plot
shows three points. In that case, the histogram consists of three peaks at
P=-0.030, 0.128 and 0.181. Thus the number of peaks in the power histogram
coincides with the multiplicity of the period. For the case of fully chaotic
motion the histogram becomes broad and continuous. We see that the histograms
can be used as a good signature for chaotic motion. Only in the case of chaotic
motion we obtain a broad, continuous distribution of the power. Since we
consider a system with a few degrees of freedom the distribution of power is
bounded from below and from above. 

To plot the distribution of power we accumulated the data into bins. For F=0.6
we have chosen the bin size to be  $\Delta=0.036$, and there are 21 bins
spanning the range (-0.378,0.378). The bin counts for a run of 11300 periods
of $\tau_{d}$ were found to be:%
\begin{eqnarray*}
\{0,0,0,0,83,305,394,677,728,648,659,795,\\
1091,1440,1108,1081,729,825,429,228,80\}.
\end{eqnarray*}
This is depicted in Figure 2.

\begin{figure}[t]
\hspace{3cm}{\epsfysize 8cm \epsfbox{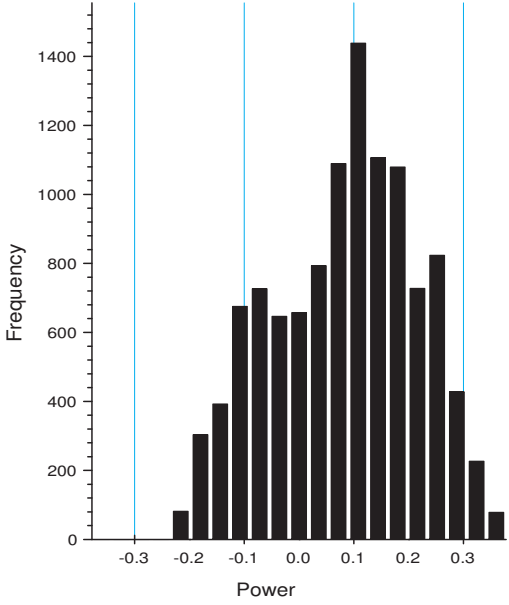}} \vspace{3mm}
\caption{Histogram of the supplied power during a period $\tau_d$}  
\label{hist1}
\end{figure}

Using the histogram we calculate the ratio of%
\begin{equation}
ratio(i)=\frac{freq(11+i)}{freq(11-i)},\ \ i=0,\cdots,6.\label{ratio}%
\end{equation}
The center of the 11'th bin corresponds to $\left\langle P\right\rangle
_{\tau}=0$. Let us define the effective temperature of the pendulum by%
\begin{equation}
\frac{1}{2}kT\equiv\frac{1}{2}\left\langle y^{2}\right\rangle _{\infty}%
=\frac{1}{2c}\left\langle P\right\rangle _{\infty},\label{temperature}%
\end{equation}
where the average is over an infinite time period. We can then define the rate
of entropy production as%
\begin{equation}
\sigma_{\tau}=\frac{\left\langle P\right\rangle _{\tau}}{kT}=\frac
{c\left\langle P\right\rangle _{\tau}}{\left\langle P\right\rangle _{\infty}%
}.\label{entropy}%
\end{equation}
Since $\left\langle P\right\rangle _{\tau}=\Delta\ i$, we can express $ratio$
as a function of $\sigma_{\tau}$ instead of $i$. In Fig. 3 we plot the
probability of having an entropy production value $\sigma_{\tau}$ during a
time interval $\tau_{d}$ vs. entropy production $-\sigma_{\tau}$. We fit this
to an exponential of the form%
\begin{equation}
\exp(\lambda\ \tau_{d}\ \sigma_{\tau}),\label{fit}%
\end{equation}
with $\lambda$ a single fitting parameter. $\lambda=1$
provides a perfect fit for the first three points (solid line).
A good fit to the first six points is
obtained with $\lambda=1.25$ (dashed line). This is quite close to the value of $\lambda=1$
predicted by the FT for a system with a large number of degrees of freedom. Of
course unlike the case for a system of many degrees of freedom the fit has to
be cut off since the power distribution is bounded (the next data point is
at infinity). Thus if one is considering
the properties of the histogram not too close to the edge, it approximately
satisfies the FT. The deviation from $\lambda=1$ seems to be
associated with the fact that the distribution is bounded due to the
small number of degrees of freedom (finite size effect).
\begin{figure}
\centerline{\epsfysize 8cm \epsfbox{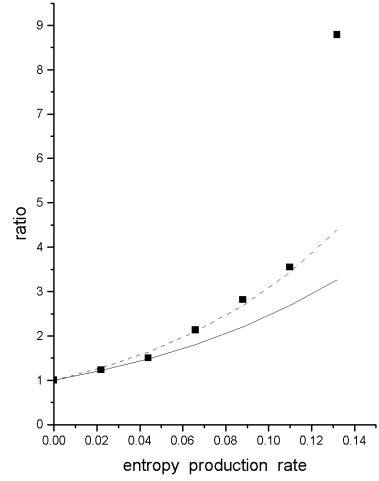}} \vspace{3mm}
\caption{Plot of $\pi(\sigma_\tau)/\pi(-\sigma_\tau)$ for
$\tau=\tau_d$.
The fit is to the form $\exp( \lambda \tau_d \sigma_\tau)$ as discussed in the text.}  
\label{fratio}
\end{figure}

We should mention that the dissipated power given by $c\left\langle
y^{2}\right\rangle _{\tau}$ also has a broad distribution for the case of
chaotic motion, but of course it is never negative. Thus unlike the
thermostatted case the two histograms of the supplied power and dissipated
power are different.

We also calculated the histogram of supplied power for a time period
$\tau=2\tau_{d}$, see Fig. 4.. The sum of frequencies is 10,000 and the bin
size is 0.027. In that case the distribution is narrower as expected, although
there are still periods of negative power. Of course in the limit of large
$\tau$ the histogram will be peaked at the average value.

\begin{figure}
\centerline{\epsfysize 8cm \epsfbox{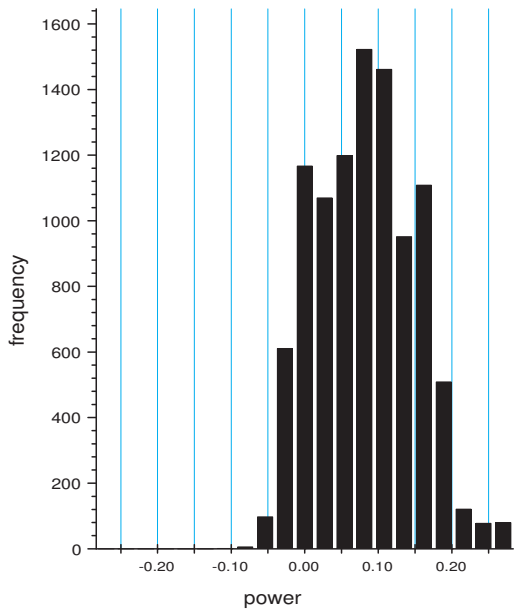}} \vspace{3mm}
\caption{Histogram of the supplied power during a period $2\tau_d$}  
\label{hist2}
\end{figure}

It is now clear that chaos is characterized by a wide, continuous distribution
of entropy production rates, whereas in the non chaotic regime the histogram
is characterized by discrete peaks whose number is equal to the multiplicity
of the period in terms of the period of the driving force. Thus the power
distribution is a good signature for the chaotic state. The supplied power can
be negative for time periods equal to the period of the driving force, but the
the distribution is centered about a positive average value. For values of the
power not too close to the edge the ratio of probabilities
\begin{equation}
\frac{\pi_{\tau_{d}}(P)}{\pi_{\tau_{d}}(-P)}\approx\exp\left(  \tau_{d}%
\frac{P}{kT}\right)  =\exp\left(  \frac{\tau_{d}c}{\left\langle P\right\rangle
_{\infty}}P\right)  .
\end{equation}
This suggests that at least near the center one may approximate the power
distribution by a gaussian of the form%
\begin{equation}
\pi_{\tau}(P)\propto\exp\left(  -\frac{\tau c}{4\left\langle P\right\rangle
_{\infty}^{2}}\left(  P-\left\langle P\right\rangle _{\infty}\right)
^{2}\right)  .
\end{equation}
It should be emphasized that a gaussian is only one possible solution
satisfying the FT, and other solutions deviating from a gaussian are
possible.

If we consider $N$ identical (uncoupled) driven dissipative pendulum, then the
total power supplied which is the sum of individual powers will have a gaussian
distribution in the limit $N\rightarrow\infty$ as predicted by the central
limit theorem. One can consider $N$ coupled pendulums, and in that case it
will be very interesting to check for the validity of the fluctuation theorem
when the motion of the system is chaotic and $N\rightarrow\infty$. In that
case the power distribution is expected to be unbounded. It need not
necessarily be gaussian. Devaitions from gaussian behavior were
observed for turbulant flows \cite{bramwell}.

Gallavotti and Cohen used the time reversal invariance of the particular
thermostatted systems that they considered in order to prove the FT. Such a
symmetry is not present in our system, although energy is still conserved when
averaged over an infinitely long time period. Of course it will be very
interesting to find a proof of the theorem for a steady state of a dissipative
chaotic system for which the energy is conserved only over long time periods
if indeed the theorem is still true.

Another ingredient of the proof of GC is the validity of the SRB measure on
the attractor. This measure implies that the relative time the system spends
at a certain region of the attractor is inversely proportional to the positive
expansion rate at that region as given by the local positive Lyapunov
exponents (for a long trajectory centered at that local). We carried out
\cite{yyg} a preliminary investigation of the attractor of the chaotic
pendulum and showed that there is a positive correlation (although not
perfect) between the time spent by the system at a region of the attractor
(divided into cells) and the inverse local expansion rate as given by the
local positive Lyapunov exponent.

This research is supported by the US Department of Energy (DOE), grant No.
DE-FG02-98ER45686. I thank Michael Widom and Itamar Procaccia for some useful discussions.


\begin{thebibliography}{9}
\bibitem{GC}G. Gallavotti and E. G. D. Cohen, Phys. Rev. Lett. \textbf{74},
2694 (1995); J. Stat Phys. \textbf{80}, 931 (1995).

\bibitem{gal} D. V. Anosov, Proc. Steklov Inst. Math. {\bf 90}, 1
(1967); G. Gallavotti, Chaos {\bf 8}, 384 (1998).

\bibitem{gallavotti}G. Gallavotti, Phys. Rev. Lett, \textbf{77}, 4334 (1996).

\bibitem{kurchan}J. Kurchan, J. Phys. A \textbf{31}, 3719 (1998).

\bibitem{LS}J. L. Lebowitz and H. Spohn, J. Stat Phys. \textbf{95}, 333
(1999).

\bibitem{york} Y-C Lai, C. Gribogi, J. A. York and I. Kan,
Nonlinearity {\bf 6}, 779 (1993). 

\bibitem{marion}J. B. Marion and S. T. Thornton in \textit{Classical dynamics
of particles and systems, }forth edition, Harcourt Brace \& Company, 1995.

\bibitem{humieres}D. D'Humieres, M.R. Beasely, B.A.Huberman and A. Libchaber,
Phys. Rev. A \textbf{26}, 3483 (1982).

\bibitem{gwinn}E.G. Gwinn and R.M. Westervelt, Phys. Rev. Lett. \textbf{54},
1613 (1985).

\bibitem{baker}G.L. Baker and J.P. Gollub in \textit{Chaotic dynamics: an
introduction} Cambridge University Press, 1990.

\bibitem{yyg}Y. Y. Goldschmidt, unpublished.

\bibitem{bramwell} S. T. Bramwell {\it et al.}, Nature {\bf 396}, 552 (1998). 
\end{thebibliography}
\end{document}